# What do we know about DNA mechanics so far?

Abhishek Aggarwal[1,a], Supriyo Naskar[1,a], Anil Kumar Sahoo[1,a],

Santosh Mogurampelly[2], Ashok Garai[3] and Prabal K. Maiti[1,*]

[1]Center for Condensed Matter Theory, Department of Physics, Indian Institute of Science, Bangalore 560012, India

[2]Department of Physics, Indian Institute of Technology Jodhpur, Rajasthan 342037, India

[3]Department of Physics, The LNM Institute of Information Technology, Sumel, Jaipur, Rajasthan 302031, India

[a] Equal Contribution

*Correspondence: maiti@iisc.ac.in

**Abstract**

The DNA molecule, apart from carrying the genetic information, plays a crucial role in a variety of biological processes and find applications in drug design, nanotechnology and nanoelectronics. The molecule undergoes significant structural transitions under the influence of forces due to physiological and non-physiological environments. Here, we summarize the insights gained from simulations and single-molecule experiments on the structural transitions and mechanics of DNA under force, as well as its elastic properties, in various environmental conditions, and discuss appealing future directions.

**Introduction**

The celebrated molecule of life, DNA is one of the most intensively studied biopolymer because of its central role in molecular and cell biology. DNA molecules are often bent, twisted, and stretched, as well as undergo various structural transitions, due to the influence of different forces arising from its biological or artificial environments[1]. DNA structural transitions play a crucial role in facilitating numerous biological functions. A deeper insight into the nature of DNA structural transitions under mechanical forces offer an excellent route to characterize the mechanical and elastic properties of DNA. Moreover, the understanding of the mechanisms of the underlying DNA structural transitions has important implications for DNA-based nano-technological applications, such as rational drug design, programmable molecular machines, and DNA origami. Due to the advent of single-molecule manipulation techniques and theoretical/computational studies, significant progress has been made in understanding the structural, mechanical and elastic properties of DNA[1-12], as well as in understanding small-molecule interactions with DNA[13-15].

In this minireview, we focus on some of the recent key developments in the field of DNA mechanics. Different theoretical models of DNA elasticity are first discussed, followed by changes in DNA elasticity in various environmental conditions, as well as DNA structural transitions, force−extension behavior, and strand separation (or melting) under external force.

**Theoretical models to study DNA mechanics**

Double-stranded DNA (dsDNA) undergoes various deformations namely twisting, bending and stretching during various transcription and genomic processes[16]. Numerous models have been proposed to understand and quantify the DNA mechanical properties[6,17-23]. The most widely used model is the Worm-like Chain (WLC) model which characterizes dsDNA as an elastic rod that bends smoothly under the random thermal fluctuations (Figure 1a). This model consists of a single fitting parameter termed as 'persistence length' and assumes quadratic form of bending deformation energies and is used extensively to calculate various elastic properties of dsDNA. The WLC model corresponds to the isotropic elastic rod model in the limit of no twist and no stretch[23]. Within the framework of elastic rod model, the energy ($E$) of a nucleic acid (NA) under the application of an external force ($F$) can be written as[24]

$$E = E_{Stretch} + E_{Twist} + E_{Twist-stretch} + E_{Bend} - force \times extension \qquad [1]$$

$$\Rightarrow E = \frac{1}{2}\frac{\gamma}{L}(\delta L)^2 + \frac{1}{2}\frac{C}{L}\varphi^2 + \frac{g}{L}(\delta L).\varphi + \frac{1}{2}\frac{\kappa}{L}\theta^2 - F.(\delta L) \qquad [2]$$

where $L$ is the equilibrium contour length, $\delta L$ is the extension in $L$, and $\varphi$ and $\theta$ are the change in helical twist and bending angle, respectively from their equilibrium values. The symbols $\gamma$, $C$, $g$ and $\kappa$ are the stretch modulus, twist modulus, twist-stretch coupling constant and bending modulus, respectively. In the absence of any external force, the Boltzmann probability distribution of contour length and bending angle can be written as[15,17,25-28]

$$P(L) = \sqrt{\frac{\gamma}{2\pi K_B TL}} exp\left[-\frac{\gamma}{2K_B TL}(\delta L)^2\right] \qquad [3]$$

$$P(\theta) = \sqrt{\frac{\kappa}{2\pi K_B T L}} exp\left[-\frac{\kappa}{2K_B T L}\theta^2\right] \qquad [4]$$

Many recent studies have implemented the above relations to compute $\gamma$, $\kappa$ and persistence length ($L_p = \frac{\kappa}{K_B T}$) of nucleic acids[15,17,25,26]. Mean square end-to-end distance ($R$) of the worm-like chain is given by[29]

$$\langle R^2 \rangle = 2L_p L_0 - 2L_p^2 \left\{1 - exp\left(-\frac{L_0}{L_p}\right)\right\} \qquad [5]$$

where $L_0$ is the contour length of the chain. This equation is a smooth crossover between ideal chain ($L_0 \gg L_p$) and a rod-like chain ($L_0 \ll L_p$). Thus, the above Eq. becomes sometime very useful to measure the persistence length of DNA[17]. Maaloum *et al.* studied the flexibility of short DNA using AFM experiments and found that the flexibility of short DNA, beyond two helical turns, can be well captured by the WLC model[28,30]. Maiti and co-workers implemented the WLC model to study the mechanics of dsDNA and DNA based nanostructures like paranemic crossover DNA molecule, DNA nanotube, etc. and found their elastic properties to be in close agreement with the experimental results (Figure 1)[31-34]. Golestanian *et al.* found that the stretch, bend and twist of the DNA depend on its length. Their study indicates that shorter DNAs are more flexible than the longer one[35]. They believe that the softening is associated with the 'cooperative' motions of the base pairs rather than its completely independent random motions. The validity of WLC model for short dsDNA molecules and at high force regimes is however challenged by some experimental studies[36]. Several modifications to WLC model[11,12,22,23] have been proposed to explain the formation of bubbles and kinks in dsDNA. The limit of the applicability of WLC model for certain length of dsDNA remains debatable[17,37]. Gross *et al.*[22] presented a twistable WLC (tWLC) model where an enthalpic twist-stretch coupling term is added to the WLC model which accounts for the changes in the DNA elongation due to variations in the twist. This description includes both the helicity and the sequence of the dsDNA and accurately explains the experimentally observed force-extension curves at all force regimes. In tWLC model, the relation between $F$ and $\delta L$ is given by[38]:

$$\delta L = L\left(1 - \frac{1}{2}\sqrt{\frac{K_BT}{F.L_p}}\right) + \frac{C}{-g^2 + \gamma C} \quad [6]$$

For WLC model, i.e., in the limit of $C \to 0$, eq. 6 reduces to $\delta L = L\left(1 - \frac{1}{2}\sqrt{\frac{K_BT}{F.L_p}}\right)$ and the $L_p$ can be obtained from the slope of curve between $\delta L$ and $F$.

Linear Sub-Elastic Chain (LSEC) model[19] was also proposed which considers linear energy function for the bending energy of dsDNA as opposed to the second order energy function in WLC model. Other than WLC model, Freely Jointed Chain (FJC) model[6] is also used to describe the dsDNA mechanics. FJC model assumes a polymer chain consisting of *n* segments of characteristic length (Kuhn length), connected via freely rotating joints[6,23] (Figure 1b). However, FJC model can only describe the behavior of dsDNA in the limit of low forces but fails to do so at intermediate and higher forces[6]. Lipfert *et al.*[20,39] proposed *springiness* model to explain the opposite twist-stretch coupling of dsRNA relative to dsDNA where dsRNA's "spring-like" properties are expected to render it more pliable to stretching. Inspired by the *springiness* model, Perez *et al.*[21] recently suggested a new parameter termed as "Crookedness" parameter to regulate the DNA mechanics at short length scales which is defined as the cosine inverse of the ratio of end-to-end length of the dsDNA to the contour length (Figure 1d). To link DNA crookedness to its stretch modulus, DNA base pairs are assumed to be beads connected via springs (Figure 1c). A low crookedness is found to be related to the high DNA stretching stiffness in their work. This model predicts the sequence-dependent elastic properties of dsDNA quite well.

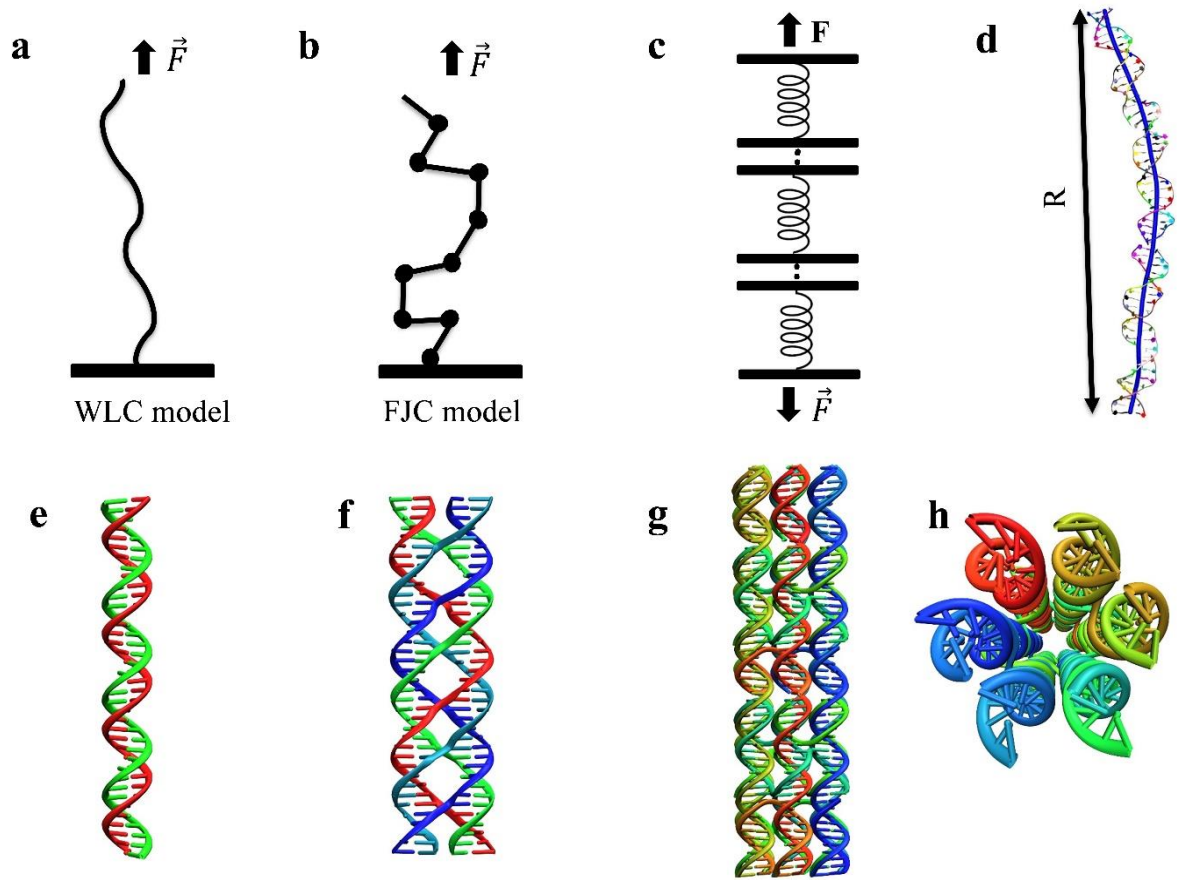

**Figure 1** Schematic diagrams of different models to measure dsDNA and DNA nanostructure's elasticity. a) Freely Jointed Chain (FJC) model, b) Worm-like Chain (WLC) model. The external force is represented by the black arrow. c) Schematic diagram to represent the dsDNA base pairs connected via springs. d) Schematic diagram to show the flexibility of long dsDNA chain. The blue line represents the central axis of dsDNA, whereas R represents the end-to-end length of the dsDNA. Clearly, the contour length of the dsDNA is very different than the sum of distances between consecutive base pairs. e)-h) Computational models of dsDNA and DNA based nanostructures where these models have been implemented. e) 38 bp dsDNA, f) Paranemic crossover DNA molecule. g) Side view and h) top view of DNA nanotube.

## DNA in Different Environments

The diverse intracellular physiological conditions surrounding DNA, in terms of ionic atmosphere, neighboring proteins and biomolecules, have a strong influence on its mechanical properties (Figure 2). Being a highly negatively charged biomolecule, the folding and compaction of DNA require overcoming an enormous electrostatic energy barrier, which is screened by the presence of positive counterions, proteins, and other biomolecules. It is intriguing how a relatively stiff and charged DNA molecule wraps around a histone octamer or forms DNA−protein complexes[40]. The mechanical properties of DNA thus change with its environment to facilitate it to make stable complexes. Of all such factors, ions influence the DNA mechanics the most in various processes like transcription, replication, DNA repair, DNA packing, and chromosome formation[41]. Recent theoretical and experimental studies reveal that with increasing cation concentration, DNA bending flexibility increases and thus its persistence length decreases. Recently, Garai *et al.* observed the opposing effect of increasing monovalent salt concentration on the bending flexibility and stretch modulus of short dsDNA[17]. Increasing salt concentration enhances the screening of DNA backbone charges, which makes dsDNA easier to bend but harder to stretch. Drozdetski *et al.* showed that multivalent ions decrease the bending rigidity of B-form dsDNA, whereas A-form dsRNA bending rigidity increases by 30%[42]. In contrast, Broekmans *et al.* found from the force-extension data that the dsDNA persistence length is insensitive to the variation in magnesium concentration[43]. Though the DNA−ion interaction is well understood, it is extremely challenging to theoretically model and explain quantitively the effect of ion interaction on DNA mechanical properties. Guilbaud *et al.*, combining Netz-Orland[44] and Trizac-Shen[45] formalism with nonlinear electrostatic effect on a charged polymer, quantitatively describe their experimental data of dsDNA persistence length over a large range of ionic concentration for various metal ions, such as $Na^+$, $K^+$, $Li^+$, $Mg^{2+}$ and $Ca^{2+}$[46]. Several other studies on DNA−ion interaction reinforce the above observations[47,48]. In contrast to the behavior exhibited by dsDNA in salt solution, both dsDNA persistence length and stretch modulus increase with ionic liquid (IL) concentration[26]. IL cations infiltrate into the grooves of dsDNA where they establish strong electrostatic and hydrophobic interactions with the nucleobases and the sugar moieties and subsequently increase dsDNA rigidity. When dsDNA comes in the vicinity of macromolecules like proteins, dendrimers, intercalators (Figure 2), the

DNA mechanics changes drastically. Binding of proteins on DNA significantly changes its elastic response[49]. Positively charged dendrimer decreases the dsDNA persistence length by neutralizing the backbone charges, whereas the stretch modulus remains unaffected[25]. In contrast, once dsDNA makes a complex with the positively charged histone octamer that partially neutralizes DNA backbone charges, both its bending and stretch modulus increase by almost four times[17]. Sahoo *et al.* recently discovered that when drug molecules bind to dsDNA in the intercalation mode, the stretch modulus increases drastically, whereas persistence length decreases similar to the scenario of dsDNA in monovalent salt solutions[15]. However, there exist conflicting reports in the literature claiming no change in DNA persistence length upon drug intercalation, as discussed in Refs. [15,50]. The mechanical properties of dsDNA in various environments are summarized in Table 1.

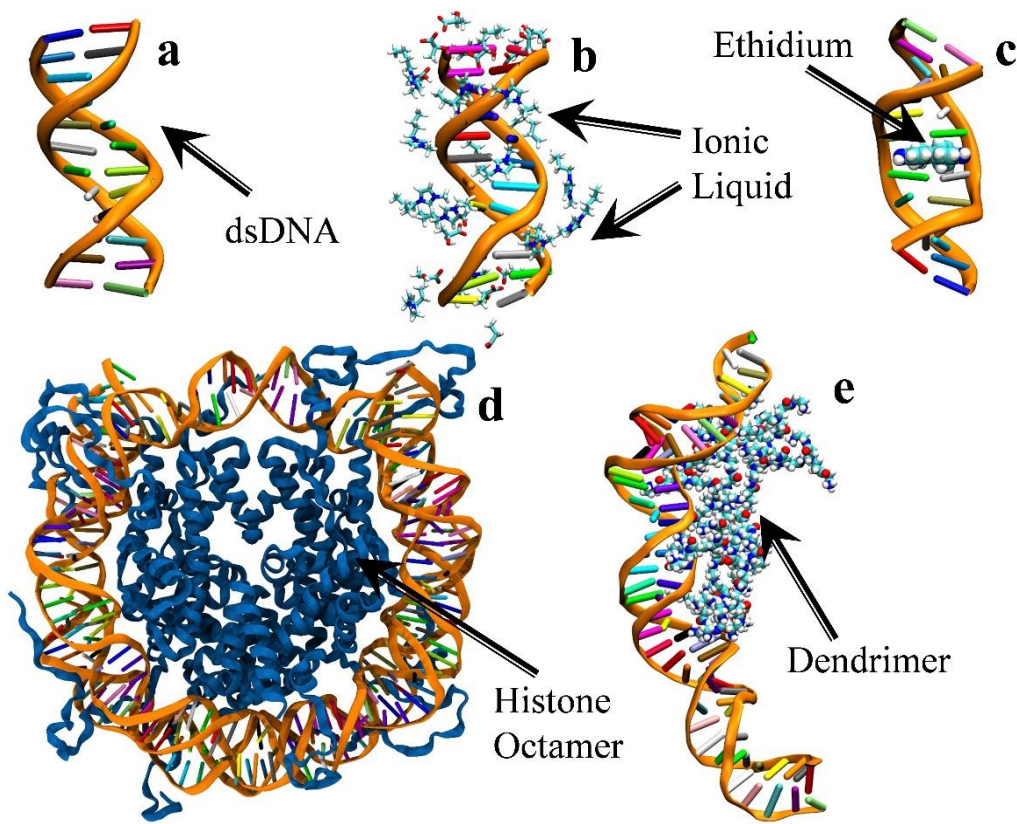

**Figure 2** Equilibrated atomistic structures of dsDNA in various environments. Molecular structures of MD simulated (a) dsDNA in aqueous conditions, (b) dsDNA in ionic liquid solution, (c) dsDNA intercalated with ethidium molecule, (d) nucleosome core particle (NCP), and (e) DNA−dendrimer complex. Water molecules are not shown here for clarity.

**Table 1.** Mechanical properties of dsDNA in various environment.

| System | Persistence Length (nm) | Stretch Modulus (pN) | Reference |
|---|---|---|---|
| **dsDNA** | 49.89 ± 0.77 | 1096 to 1280 | [17,24] |
| **dsDNA in 250mM NaCl** | 33.16 ± 0.45 | 2345 ± 65 | [17] |
| **NCP DNA** | 59.02 ± 6.24 | 4021 ± 151 | [17] |
| **ethidium intercalated dsDNA** | 30.97 ± 1.77 | 2118.6 ± 102.3 | [15] |
| **dsDNA–dendrimer complex** | 6.3 | 959 | [25] |
| **dsDNA in 1-butyl-3-methyl imidazolium acetate** | 248.74 ± 6.8 | 3926.24 ± 124.02 | [26] |
| **Bare dsDNA in Physiological Condition (experiment)** | 39 to 50 | 1450 to 1750 | [38,43] |

**Structural transitions in dsDNA under force**

dsDNA exhibits different force−extension regimes with Hookean behavior at low forces and a drastic elongation of its length at a force of 65 pN[1,6]. This is followed by an overstretching behavior where dsDNA transforms to a completely new structural phase − broadly described as a thermodynamically melted form (with broken Watson-Crick base-pairs) or a structurally intact base-pair stacking alignment in ladder-like fashion (the so-called S-DNA)[6,51-53].

In the force induced melting perspective, Bloomfield *et al.*[8] proposed a theoretical model based on the underlying thermodynamics to explain quantitatively the experimentally observed overstretching transition. Dynamics of DNA under external force in a coarse grain level can be found in a review by Kumar and Li[54]. Computer simulations using atomistic models by Maiti and co-workers[4,5,15,33,51,52,55] expanded the understanding of the nature of underlying mechanisms with microscopic details and provided deeper insights into the force induced DNA melting description of the DNA structural transitions. Besides, several researchers reported the observation of S-DNA, under specific force attachment at the end points of strands by using both single-molecule experiments and computer simulations[5,7,9,10] (Figure 3).

While the nature of DNA structural phase transitions under applied forces has been well understood on a broad perspective, there are few challenges in expanding the application spectrum of DNA-based materials[32,33,56]. Firstly, there is a lack of fundamental understanding of the dynamics of DNA undergoing overstretching transition due to interaction with proteins and other biomolecules, the orientation of base-pair tilt and its sensitivity with respect to the presence of biomacromolecules etc. Despite the requirement for a thorough examination of the above subject in the context of DNA role in rational drug design applications, there were only a few preliminary reports dealing with the above-mentioned issues. Mameren *et al.*[57] investigated the molecular level dynamic features exhibited by intercalated dsDNA under a large range of pulling forces using polarized fluorescence microscopy in conjunction with optical-tweezers-based single-molecule manipulation techniques (Figure 3a). They measured the alignment of the transition dipoles of the intercalating dyes when the DNA is pulled tightly and observed that on average the intercalated dyes are oriented perpendicular to the long axis of the DNA. Moreover, the dyes undergo fast dynamics on the time scales of absorption and fluorescence emission processes. It is also reported that the intercalating dyes remain bounded to DNA even beyond overstretching forces where DNA is melted. Peterman, *et al.*[58] used single-molecule polarization microscopy to examine the structure of S-DNA with the help of intercalated dyes in the regime of DNA under tension. In contrast to the perpendicular orientations typically observed at low stretching forces, they find that the intercalator dyes adopt a tilted orientation of 54° relative to the DNA long axis at stretching forces beyond the dramatic overstretching transition. It was concluded that the S-DNA has substantially inclined base pairs relative to those of the canonical B-DNA conformation. Note that the novel approach devised by Peterman *et al.*[58] is a significant achievement in determining the inclination of intercalated dye molecules. However, it is not possible to directly measure the orientations of the DNA base-pairs in overstretched regime and therefore it is hard to examine the effect of intercalators on the base-pair inclination of overstretched form of DNA (specifically in this context S-DNA). Further efforts are required to establish a complete picture of structural basis for the DNA conformational changes that enable genomic transactions and biotechnological processing of DNA. It is in this context the drug design applications based on DNA require more serious efforts.

**Force-induced DNA strand separation**

In various biological processes, such as DNA transcription (with the help of RNA polymerases), replication (with the help of helicases), and repair, the two strands of a dsDNA are melted from the middle or ends of the DNA, either partially or fully[1]. The design principle of a variety of drug molecules, including that for cancer and tumour treatments, is based on the principle that these molecules bind to DNA and make dsDNA melting extremely difficult, which thus could hamper the biological processes required for proper functioning of the cellular processes. In this context, Sahoo *et al.* [15] employed steered molecular dynamics (MD) simulations of different pulling protocols to study the overstretching and melting transitions of dsDNA and found a significant increase in the dsDNA rupture force in the presence of intercalated drug molecules, which is due to the enhancement in the base-stacking interaction between adjacent base pairs present near to a drug intercalation site (Figure 3). This is in line with the recent revelation that the base-stacking interactions, instead of the previously thought base-pairing interactions, mostly contribute to the stability of dsDNA[59]. Moreover, they found that in the presence of intercalated drug molecules, dsDNA stretch modulus increases significantly, in agreement with several experimental reports[60,61]. Their study revealed that anticancer drugs could function by altering the mechanical properties of DNA.

DNA methylation (or hydroxymethylation) is known to play a vital role in epigenetic regulation of gene expression. A combined experimental and simulation study[62] revealed that the rupture force for DNA strand separation increases upon methylation of its Cytosine bases due to the enhancement in base-stacking interaction. They also showed that the rupture force not only depends on the methylation level but also on the sequences of methylated bases. The same but more pronounced effects were observed for Cytosine hydroxymethylation[55]. These studies thus provided a mechanism that methylation or hydroxymethylation, via altering DNA mechanical properties, could control biological functions.

Many DNA-associated biological processes hinge on sequence-selective molecular interactions[63]. By using the force-induced remnant magnetization spectroscopy technique, Hu *et al.*[63] measured the differential binding force to demonstrate the DNA sequence-specific binding of daunomycin (an anticancer drug), as well as to reveal the enantioselective binding of tetrahydropalmatine (L or D form) with different DNA sequences—the understanding of which has important implications for pharmaceutical research. Manosas *et al.*[64] used an assay of mechanical unzipping of DNA hairpins to determine sequence-selective binding of small ligands. This study revealed that actinomycin (a mono-intercalator) likes to bind GpC steps, whereas bis-intercalators thiocoraline and echinomycin like to bind CpG steps.

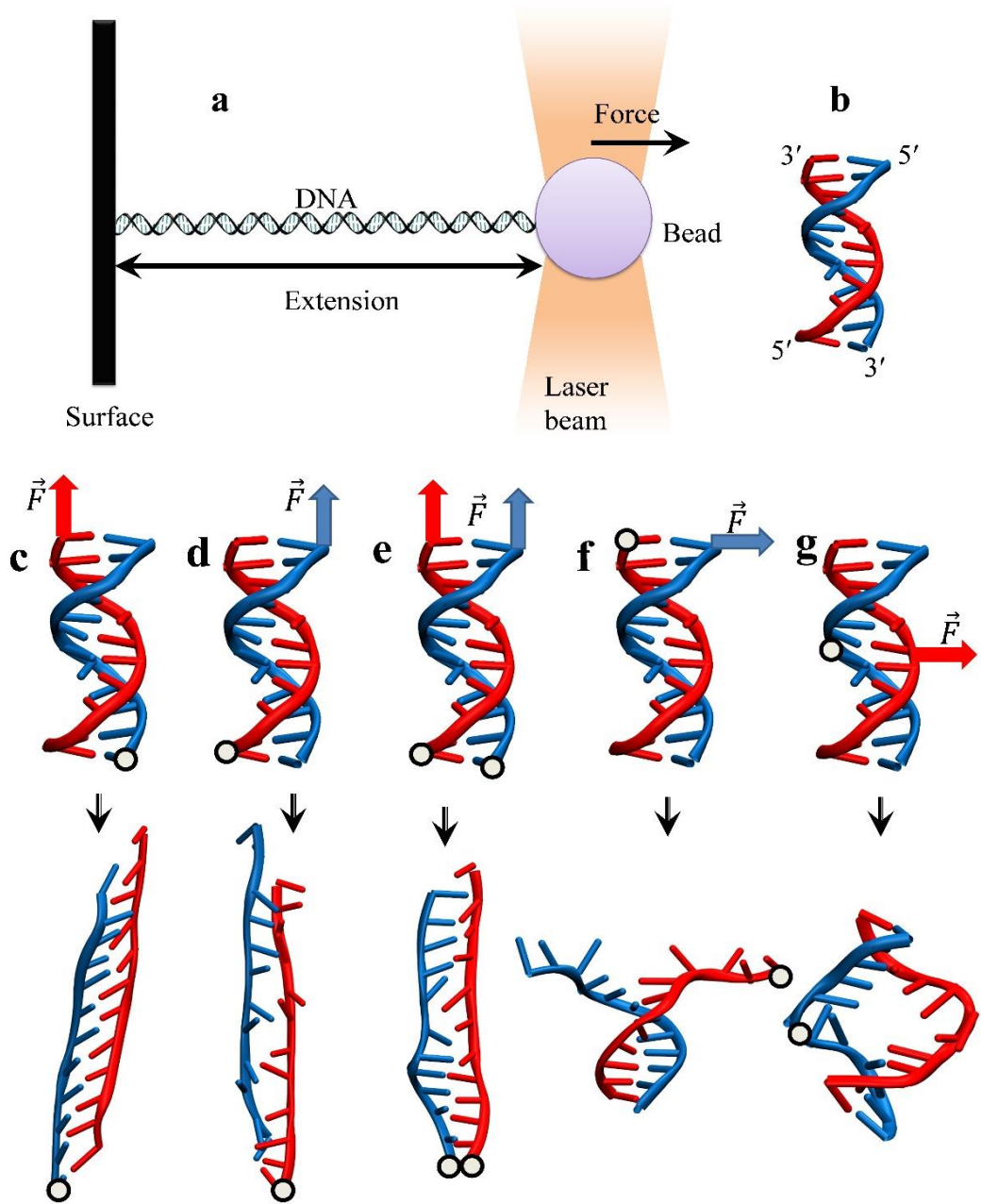

**Figure 3** Pulling protocol dependent structural transitions of dsDNA. a) Schematic diagram to represent the experimental setup to stretch dsDNA using optical tweezers. b) Schematic diagram to highlight the 3' and 5' ends of dsDNA. c) – g) Stretching dsDNA using various pulling protocols. Here, the black circle represents the fixed end of dsDNA while red and blue arrows represent the direction of force applied to dsDNA along c) 3' end, d) 5' end, e) both 3' and 5' ends in the same side, f) unzipping force applied at terminal base of dsDNA and g) unzipping force applied at the middle of dsDNA. Stretching dsDNA along 3' end leads to a ladder-like S-form of DNA, whereas stretching along 5' end leads to melting of dsDNA. Stretching a dsDNA along both 3' and 5' ends leads to an intermediate state where some portions of DNA show ordered base pairs, while others show melting. Unzipping dsDNA along ends and middle of dsDNA leads to its tearing in different fashions.

**Conclusions and Future Perspectives**

Advances in single-molecule experiment techniques and insights from computational simulations have greatly expanded our understanding of DNA mechanics, DNA-associated cellular processes, and functioning of various drug molecules that target DNA. dsDNA has been studied in various conditions, such as unzipping, shearing, stretching, and under a torque, to mimic different biological scenarios[1]. These have provided increased insights to various interesting aspects of DNA such as emergence of S-DNA upon overstretching[5,10,51,52], modifications of DNA structural and mechanical properties under different environments such as drug intercalators[15], salts[17,46] and ionic liquids[26]. These information are the key inputs for DNA nanotechnological applications, enabling researchers to design various DNA nanostructures like DNA nanotubes[32,33] and DNA nanorobots for drug delivery[65], as well as in DNA nanoelectronics[51,52].

Use of DNA for nanobiotechnological applications still requires a complete understanding of its mechanical properties at different length scales and in complex and crowded environments. Further investigations, combining experiment, simulation, and theory, are required to fully understand the dependence of DNA persistence length on its contour length and sequence, as well as its interaction with salts, proteins, and small-molecule ligands. How confinement affects DNA structural and elastic properties will be an important direction to pursue from theoretical/simulation point of view, as it has high relevance for cell biology, it is less studied and understood, and probing it experimentally is a rather difficult task. Moreover, the structural transitions of DNA in confined geometry, under the application of an external force can be obtained with unprecedented resolution from molecular simulations.

**Conflict of interest statement**

None declared.

**Acknowledgements**

A.A., S.N., and A.K.S. thank MHRD India, CSIR India, and IISc, respectively for the generous fellowship. We thank DAE, India for financial support. AG thanks SERB, DST, Govt. of India for the financial support (grant no ECR/2017/000683).